\documentclass[12pt]{article}

\usepackage{amsmath}
\usepackage{stmaryrd}
\usepackage{subfigure}
\usepackage{empheq}
\usepackage{appendix,epic,eepic,amsmath,amsthm,amssymb,epsfig,cite,indentfirst,oldgerm}

\newcommand{\EQ}{\begin{equation}}
\newcommand{\EN}{\end{equation}}

\newcommand{\bear}{\begin{eqnarray}}
\newcommand{\ear}{\end{eqnarray}}
\newcommand{\bt} { \begin{tabular} }
\newcommand{\et}{ \end{tabular} }
\newcommand{\bc} { \begin{center} }
\newcommand{\ec}{ \end{center} }

\newcommand{\btb} { \begin{table} }
\newcommand{\etb}{ \end{table} }

\begin{document}

\topmargin 0pt
\oddsidemargin 5mm
\newcommand{\NP}[1]{Nucl.\ Phys.\ {\bf #1}}
\newcommand{\PL}[1]{Phys.\ Lett.\ {\bf #1}}
\newcommand{\NC}[1]{Nuovo Cimento {\bf #1}}
\newcommand{\CMP}[1]{Comm.\ Math.\ Phys.\ {\bf #1}}
\newcommand{\PR}[1]{Phys.\ Rev.\ {\bf #1}}
\newcommand{\PRL}[1]{Phys.\ Rev.\ Lett.\ {\bf #1}}
\newcommand{\MPL}[1]{Mod.\ Phys.\ Lett.\ {\bf #1}}
\newcommand{\JETP}[1]{Sov.\ Phys.\ JETP {\bf #1}}
\newcommand{\TMP}[1]{Teor.\ Mat.\ Fiz.\ {\bf #1}}

\renewcommand{\thefootnote}{\fnsymbol{footnote}}

\newpage
\setcounter{page}{0}
\begin{titlepage}
\begin{flushright}

\end{flushright}
\vspace{0.5cm}
\begin{center}
{\large Integrability of the odd eight-vertex vertex model with symmetric weights} \\
\vspace{1cm}
{\large M.J. Martins } \\
\vspace{0.15cm}
{\em Universidade Federal de S\~ao Carlos\\
Departamento de F\'{\i}sica \\
C.P. 676, 13565-905, S\~ao Carlos (SP), Brazil\\}
\vspace{0.35cm}
\end{center}
\vspace{0.5cm}
\vspace{0.5cm}
\begin{abstract}
In this paper we investigate the integrability properties of a two-state vertex model
on the square lattice
whose microstates at a vertex has always an odd number of incoming or outcoming arrows. 
This model was named odd eight-vertex model by Wu and Kunz \cite{WK} to distinguish 
it from the
well known eight-vertex model possessing an even number of arrows 
orientations at each vertex.
When the energy weights are invariant under arrows inversion we show that 
the integrable manifold of the odd eight-vertex model coincides with that of
the even eight-vertex model. The form of the $\mathrm{R}$-matrix 
for the odd eight-vertex
model is however not the same as that of the respective Lax operator.
Altogether we find that these eight-vertex models 
give rise to a generic sheaf of
$\mathrm{R}$-matrices satisfying the Yang-Baxter equations resembling
intertwiner relations associated to equidimensional representations.

\end{abstract}

\vspace{.15cm} \centerline{}
\vspace{.1cm} \centerline{Keywords: eight-vertex models, Yang-Baxter equations }
\vspace{.15cm} \centerline{November~~2017}

\end{titlepage}


\pagestyle{empty}

\newpage

\pagestyle{plain}
\pagenumbering{arabic}

\renewcommand{\thefootnote}{\arabic{footnote}}
\newtheorem{proposition}{Proposition}
\newtheorem{pr}{Proposition}
\newtheorem{remark}{Remark}
\newtheorem{re}{Remark}
\newtheorem{theorem}{Theorem}
\newtheorem{theo}{Theorem}

\def\ll{\left\lgroup}
\def\rr{\right\rgroup}

\newtheorem{Theorem}{Theorem}[section]
\newtheorem{Corollary}[Theorem]{Corollary}
\newtheorem{Proposition}[Theorem]{Proposition}
\newtheorem{Conjecture}[Theorem]{Conjecture}
\newtheorem{Lemma}[Theorem]{Lemma}
\newtheorem{Example}[Theorem]{Example}
\newtheorem{Note}[Theorem]{Note}
\newtheorem{Definition}[Theorem]{Definition}

\section{Introduction}

The name vertex model is used to denote a lattice model 
in which the
statistical configurations sit on each line connecting a pair 
of nearest 
neighbor sites of the lattice. These type of models first 
emerged in the
discussion by Pauling of the residual entropy 
of ice \cite{PA} and 
afterwards in the study of certain phase 
transition exhibited by
hydrogen-bonded crystals \cite{SLA}. 
In these situations the microstates
are characterized by two possible positions 
of the hydrogens commonly represented by incoming 
and outcoming
arrows placed along 
the lattice links \cite{LW}. It turns out that on the 
square lattice we have 
sixteen vertex possibilities and  
these 
configurations can be
organized in terms of two distinct families 
of eight-vertex states according to the even
or odd number arrows orientations at a vertex. 
When the number
of in and out arrows are even we 
have the standard
eight-vertex model whose vertices are shown
in Figure (\ref{figure1}). The corresponding
configurations with 
odd number of arrows 
are exhibited in Figure (\ref{figure2}).
\setlength{\unitlength}{2500sp}
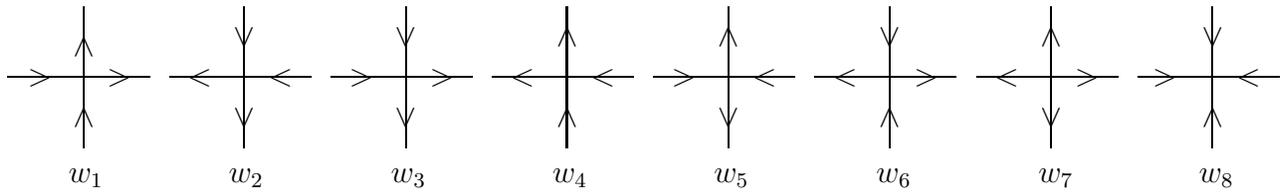
\begin{figure}[ht]
\begin{center}
\begin{picture}(8000,2000)
{\put(-2200,900){\line(1,0){1400}}}
{\put(-600,900){\line(1,0){1400}}}
{\put(1000,900){\line(1,0){1400}}}
{\put(2600,900){\line(1,0){1400}}}
{\put(4200,900){\line(1,0){1400}}}
{\put(5800,900){\line(1,0){1400}}}
{\put(7400,900){\line(1,0){1400}}}
{\put(9000,900){\line(1,0){1400}}}
{\put(-1450,1600){\line(0,-1){1400}}}
{\put(138,1600){\line(0,-1){1400}}}
{\put(1740,1600){\line(0,-1){1400}}}
{\put(3338,1600){\line(0,-1){1400}}}
{\put(4940,1600){\line(0,-1){1400}}}
{\put(6540,1600){\line(0,-1){1400}}}
{\put(8140,1600){\line(0,-1){1400}}}
{\put(9740,1600){\line(0,-1){1400}}}
{\put(-1885,890){\makebox(0,0){\fontsize{12}{14}\selectfont $>$}}}
{\put(-300,890){\makebox(0,0){\fontsize{12}{14}\selectfont $<$}}}
{\put(1300,890){\makebox(0,0){\fontsize{12}{14}\selectfont $>$}}}
{\put(2900,890){\makebox(0,0){\fontsize{12}{14}\selectfont $<$}}}
{\put(4500,890){\makebox(0,0){\fontsize{12}{14}\selectfont $>$}}}
{\put(6100,890){\makebox(0,0){\fontsize{12}{14}\selectfont $<$}}}
{\put(7700,890){\makebox(0,0){\fontsize{12}{14}\selectfont $<$}}}
{\put(9265,890){\makebox(0,0){\fontsize{12}{14}\selectfont $>$}}}
{\put(-1100,890){\makebox(0,0){\fontsize{12}{14}\selectfont $>$}}}
{\put(500,890){\makebox(0,0){\fontsize{12}{14}\selectfont $<$}}}
{\put(2100,890){\makebox(0,0){\fontsize{12}{14}\selectfont $>$}}}
{\put(3700,890){\makebox(0,0){\fontsize{12}{14}\selectfont $<$}}}
{\put(5300,890){\makebox(0,0){\fontsize{12}{14}\selectfont $<$}}}
{\put(6900,890){\makebox(0,0){\fontsize{12}{14}\selectfont $>$}}}
{\put(8500,890){\makebox(0,0){\fontsize{12}{14}\selectfont $>$}}}
{\put(10100,890){\makebox(0,0){\fontsize{12}{14}\selectfont $<$}}}
{\put(-1450,1200){\makebox(0,0){\fontsize{12}{14}\selectfont $\wedge$}}}
{\put(140,1300){\makebox(0,0){\fontsize{12}{14}\selectfont $\vee$}}}
{\put(1740,1300){\makebox(0,0){\fontsize{12}{14}\selectfont $\vee$}}}
{\put(3340,1300){\makebox(0,0){\fontsize{12}{14}\selectfont $\wedge$}}}
{\put(4940,1300){\makebox(0,0){\fontsize{12}{14}\selectfont $\wedge$}}}
{\put(6540,1300){\makebox(0,0){\fontsize{12}{14}\selectfont $\vee$}}}
{\put(8140,1300){\makebox(0,0){\fontsize{12}{14}\selectfont $\wedge$}}}
{\put(9740,1300){\makebox(0,0){\fontsize{12}{14}\selectfont $\vee$}}}
{\put(-1450,500){\makebox(0,0){\fontsize{12}{14}\selectfont $\wedge$}}}
{\put(140,500){\makebox(0,0){\fontsize{12}{14}\selectfont $\vee$}}}
{\put(1740,500){\makebox(0,0){\fontsize{12}{14}\selectfont $\vee$}}}
{\put(3340,500){\makebox(0,0){\fontsize{12}{14}\selectfont $\wedge$}}}
{\put(4940,500){\makebox(0,0){\fontsize{12}{14}\selectfont $\vee$}}}
{\put(6540,500){\makebox(0,0){\fontsize{12}{14}\selectfont $\wedge$}}}
{\put(8140,500){\makebox(0,0){\fontsize{12}{14}\selectfont $\vee$}}}
{\put(9740,500){\makebox(0,0){\fontsize{12}{14}\selectfont $\wedge$}}}
{\put(-1430,-100){\makebox(0,0){\fontsize{12}{14}\selectfont $w_1$}}}
{\put(160,-100){\makebox(0,0){\fontsize{12}{14}\selectfont $w_2$}}}
{\put(1770,-100){\makebox(0,0){\fontsize{12}{14}\selectfont $w_3$}}}
{\put(3370,-100){\makebox(0,0){\fontsize{12}{14}\selectfont $w_4$}}}
{\put(4970,-100){\makebox(0,0){\fontsize{12}{14}\selectfont $w_5$}}}
{\put(6580,-100){\makebox(0,0){\fontsize{12}{14}\selectfont $w_6$}}}
{\put(8190,-100){\makebox(0,0){\fontsize{12}{14}\selectfont $w_7$}}}
{\put(9780,-100){\makebox(0,0){\fontsize{12}{14}\selectfont $w_8$}}}
\end{picture}
\end{center}
\caption{The vertices configurations of the {\bf even}  eight vertex model 
with the respective energy weights. }
\label{figure1}
\end{figure}
\setlength{\unitlength}{2500sp}
\begin{figure}[ht]
\begin{center}
\begin{picture}(8000,2000)
{\put(-2200,900){\line(1,0){1400}}}
{\put(-600,900){\line(1,0){1400}}}
{\put(1000,900){\line(1,0){1400}}}
{\put(2600,900){\line(1,0){1400}}}
{\put(4200,900){\line(1,0){1400}}}
{\put(5800,900){\line(1,0){1400}}}
{\put(7400,900){\line(1,0){1400}}}
{\put(9000,900){\line(1,0){1400}}}
{\put(-1450,1600){\line(0,-1){1400}}}
{\put(138,1600){\line(0,-1){1400}}}
{\put(1740,1600){\line(0,-1){1400}}}
{\put(3338,1600){\line(0,-1){1400}}}
{\put(4940,1600){\line(0,-1){1400}}}
{\put(6540,1600){\line(0,-1){1400}}}
{\put(8140,1600){\line(0,-1){1400}}}
{\put(9740,1600){\line(0,-1){1400}}}
{\put(-1885,890){\makebox(0,0){\fontsize{12}{14}\selectfont $>$}}}
{\put(-300,890){\makebox(0,0){\fontsize{12}{14}\selectfont $<$}}}
{\put(1300,890){\makebox(0,0){\fontsize{12}{14}\selectfont $>$}}}
{\put(2900,890){\makebox(0,0){\fontsize{12}{14}\selectfont $<$}}}
{\put(4500,890){\makebox(0,0){\fontsize{12}{14}\selectfont $>$}}}
{\put(6100,890){\makebox(0,0){\fontsize{12}{14}\selectfont $<$}}}
{\put(7700,890){\makebox(0,0){\fontsize{12}{14}\selectfont $<$}}}
{\put(9265,890){\makebox(0,0){\fontsize{12}{14}\selectfont $>$}}}
{\put(-1100,890){\makebox(0,0){\fontsize{12}{14}\selectfont $>$}}}
{\put(500,890){\makebox(0,0){\fontsize{12}{14}\selectfont $<$}}}
{\put(2100,890){\makebox(0,0){\fontsize{12}{14}\selectfont $>$}}}
{\put(3700,890){\makebox(0,0){\fontsize{12}{14}\selectfont $<$}}}
{\put(5300,890){\makebox(0,0){\fontsize{12}{14}\selectfont $<$}}}
{\put(6900,890){\makebox(0,0){\fontsize{12}{14}\selectfont $>$}}}
{\put(8500,890){\makebox(0,0){\fontsize{12}{14}\selectfont $>$}}}
{\put(10100,890){\makebox(0,0){\fontsize{12}{14}\selectfont $<$}}}
{\put(-1450,1200){\makebox(0,0){\fontsize{12}{14}\selectfont $\wedge$}}}
{\put(140,1300){\makebox(0,0){\fontsize{12}{14}\selectfont $\vee$}}}
{\put(1740,1300){\makebox(0,0){\fontsize{12}{14}\selectfont $\vee$}}}
{\put(3340,1300){\makebox(0,0){\fontsize{12}{14}\selectfont $\wedge$}}}
{\put(4940,1300){\makebox(0,0){\fontsize{12}{14}\selectfont $\wedge$}}}
{\put(6540,1300){\makebox(0,0){\fontsize{12}{14}\selectfont $\vee$}}}
{\put(8140,1300){\makebox(0,0){\fontsize{12}{14}\selectfont $\wedge$}}}
{\put(9740,1300){\makebox(0,0){\fontsize{12}{14}\selectfont $\vee$}}}
{\put(-1450,500){\makebox(0,0){\fontsize{12}{14}\selectfont $\vee$}}}
{\put(140,500){\makebox(0,0){\fontsize{12}{14}\selectfont $\wedge$}}}
{\put(1740,500){\makebox(0,0){\fontsize{12}{14}\selectfont $\wedge$}}}
{\put(3340,500){\makebox(0,0){\fontsize{12}{14}\selectfont $\vee$}}}
{\put(4940,500){\makebox(0,0){\fontsize{12}{14}\selectfont $\wedge$}}}
{\put(6540,500){\makebox(0,0){\fontsize{12}{14}\selectfont $\vee$}}}
{\put(8140,500){\makebox(0,0){\fontsize{12}{14}\selectfont $\wedge$}}}
{\put(9740,500){\makebox(0,0){\fontsize{12}{14}\selectfont $\vee$}}}
{\put(-1430,-100){\makebox(0,0){\fontsize{12}{14}\selectfont $v_1 $}}}
{\put(160,-100){\makebox(0,0){\fontsize{12}{14}\selectfont $ v_2 $}}}
{\put(1770,-100){\makebox(0,0){\fontsize{12}{14}\selectfont $v_3 $}}}
{\put(3370,-100){\makebox(0,0){\fontsize{12}{14}\selectfont $v_4$}}}
{\put(4970,-100){\makebox(0,0){\fontsize{12}{14}\selectfont $v_5 $}}}
{\put(6580,-100){\makebox(0,0){\fontsize{12}{14}\selectfont $v_6 $}}}
{\put(8185,-100){\makebox(0,0){\fontsize{12}{14}\selectfont $v_7 $}}}
{\put(9775,-100){\makebox(0,0){\fontsize{12}{14}\selectfont $v_8 $}}}
\end{picture}
\end{center}
\caption{The vertices configurations of the {\bf odd} eight vertex model 
with the respective energy weights.}
\label{figure2}
\end{figure}
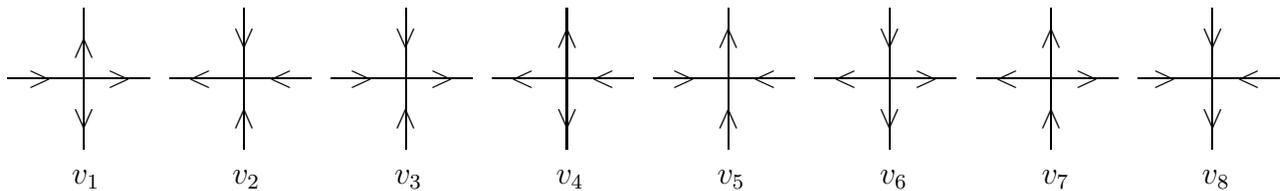

The even eight-vertex model is known to have an integrable manifold when
the weights are invariant under inversion of all arrows 
at each vertex \cite{BAX}. In this
situation the energy weights are symmetric,
\EQ
\label{IDEN}
w_1=w_2=a,~~w_3=w_4=b,~~w_5=w_6=c,~~w_7=w_8=d,
\EN

In this case 
Baxter \cite{BAX} found that there exists
a family of commuting transfer matrices provide 
the weights are 
lying on the following
intersection of quadrics,
\begin{equation}
\label{MANI}
\frac{ab-cd}{ab+cd}=\Gamma~~~\mathrm{and}~~~\frac{a^2+b^2-c^2-d^2}{2(ab+cd)}=\Delta,
\end{equation}
where $\Gamma$ and $\Delta$ are free parameters.

A natural question to be asked is whether or not one can also build 
commuting transfer matrices for
the odd eight-vertex 
model on the subspace of
symmetric weights, namely
\EQ
\label{IDEN1}
v_1=v_2={\bf a},~~v_3=v_4={\bf b},~~v_5=v_6={\bf c},~~v_7=v_8={\bf d},
\EN
and what is the respective algebraic variety constraining the weights ${\bf a,b,c}$ and ${\bf d}$.

In this paper we shall argue that both even and odd 
symmetric eight-vertex models are integrable when their weights
are lying on the same algebraic curve (\ref{MANI}). Altogether they
give rise to a sheaf of $\mathrm{R}$-matrices satisfying 
the Yang-Baxter equations with structure similar to the
intertwiner relations of two distinct representations
of a given algebra. The common invariance here is the symmetry of
the weights upon the inversion of the arrows at a vertex.

We start next section recalling a result by Wu and Kunz \cite{WK}
who have shown that 
the partition function of the odd eight-vertex model can be expressed
as the partition function 
of an alternating bipartite even eight-vertex model. This equivalence
can be reversed and the partition function of even eight-vertex model 
can also be written as that of a staggered odd eight-vertex model.
These mappings for symmetric weights suggest that both even
and odd eight-vertex may be integrable on the same algebraic
manifold. This is elaborated in section $3$ where we show that  
the transfer matrices of the even and odd eight-vertex models
with symmetric weights are in fact related. 
We then study the Yang-Baxter equation for the symmetric odd eight-vertex
uncovering the structure of the respective $\mathrm{R}$-matrix which turns
out to be distinct from that of the Lax operator. This guided us
to propose the form of four types of $\mathrm{R}$-matrices satisfying 
a set of Yang-Baxter relations on the same elliptic curve. In section 4
we comment on the possibility of integrability in the case of asymmetric weights.

\section{Staggered eight-vertex models}

In the staggered eight-vertex model it is allowed
different vertex weights for the two possible sublattices 
of a $2\mathrm{N} \times 2\mathrm{N}$
square lattice. 
This is schematized in Figure (\ref{figure3}) in which
the sublattices are represented by the symbols $\mathrm{X}$ and $\mathrm{Y}$. For
the sublattice $\mathrm{X}$ we assign the weights $\omega_{\mathrm{X}}$ while for
the sublattice $\mathrm{Y}$ the weights are
$\omega_{\mathrm{Y}}$. 
\setlength{\unitlength}{3500sp}
\begin{figure}[ht]
\begin{center}
\begin{picture}(6537,3912)(2000,-6361)
\put(3900,-6400){\makebox(0,0){\fontsize{10}{12}\selectfont 1}}
\put(4500,-6400){\makebox(0,0){\fontsize{10}{12}\selectfont 2}}
\put(5140,-6400){\makebox(0,0){\fontsize{10}{12}\selectfont 3 }}
\put(6300,-6400){\makebox(0,0){\fontsize{10}{12}\selectfont 2N}}
\put(5701,-6400){\makebox(0,0){$\dots$}}
\put(3260,-5760){\makebox(0,0){\fontsize{10}{12}\selectfont 1}}
\put(3260,-5170){\makebox(0,0){\fontsize{10}{12}\selectfont 2}}
\put(3260,-4570){\makebox(0,0){\fontsize{10}{12}\selectfont 3}}
\put(3260,-3980){\makebox(0,0){\fontsize{10}{12}\selectfont }}
\put(3260,-3360){\makebox(0,0){\fontsize{10}{12}\selectfont 2N }}
\put(3260,-3986){\makebox(0,0){$\vdots$}}
\put(3401,-3361){\line( 1, 0){3500}}
\put(3401,-4561){\line( 1, 0){3400}}
\put(3401,-5161){\line( 1, 0){3400}}
\put(3401,-5761){\line( 1, 0){3400}}
\put(3901,-6261){\line( 0, 1){3400}}
\put(4501,-6261){\line( 0, 1){3400}}
\put(5101,-6261){\line( 0, 1){3400}}
\put(6301,-6261){\line( 0, 1){3400}}
\put(3785,-5680){\makebox(0,0){\fontsize{10}{12}\selectfont X}}
\put(4385,-5680){\makebox(0,0){\fontsize{10}{12}\selectfont Y}}
\put(4985,-5680){\makebox(0,0){\fontsize{10}{12}\selectfont X}}
\put(6195,-5680){\makebox(0,0){\fontsize{10}{12}\selectfont Y}}
\put(3785,-5080){\makebox(0,0){\fontsize{10}{12}\selectfont Y}}
\put(4385,-5080){\makebox(0,0){\fontsize{10}{12}\selectfont X}}
\put(4985,-5080){\makebox(0,0){\fontsize{10}{12}\selectfont Y}}
\put(6195,-5080){\makebox(0,0){\fontsize{10}{12}\selectfont X}}
\put(3785,-4480){\makebox(0,0){\fontsize{10}{12}\selectfont X}}
\put(4385,-4480){\makebox(0,0){\fontsize{10}{12}\selectfont Y}}
\put(4985,-4480){\makebox(0,0){\fontsize{10}{12}\selectfont X}}
\put(6195,-4480){\makebox(0,0){\fontsize{10}{12}\selectfont Y}}
\put(3785,-3280){\makebox(0,0){\fontsize{10}{12}\selectfont Y}}
\put(4385,-3280){\makebox(0,0){\fontsize{10}{12}\selectfont X}}
\put(4985,-3280){\makebox(0,0){\fontsize{10}{12}\selectfont Y}}
\put(6195,-3280){\makebox(0,0){\fontsize{10}{12}\selectfont X}}
\end{picture} \par
\end{center}
\caption{The staggered eight-vertex model with 
vertex weights $\omega_{\mathrm{X}}$ on sublattice $\mathrm{X}$ and
$\omega_{\mathrm{Y}}$ on  sublattice $\mathrm{Y}$.
\label{figure3}}
\end{figure}
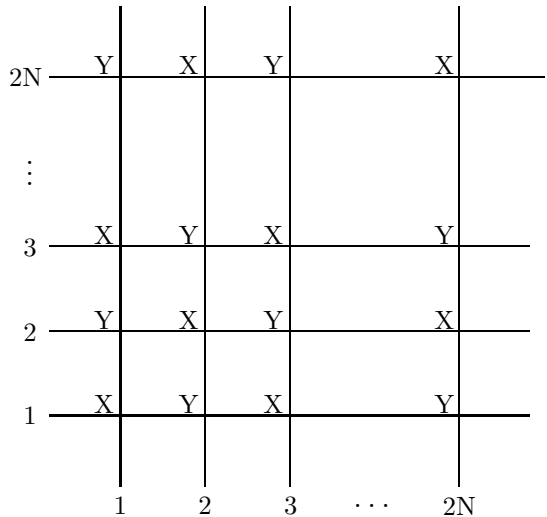

Let us denote by $\mathrm{Z}_{\mathrm{8\mathrm{ev}}}^{(\mathrm{stag})}(w_{\mathrm{X}};w_{\mathrm{Y}})$
the partition function of the staggered model built using the weights configurations of the
even eight-vertex model. From the work by Wu and Kunz \cite{WK}
one concludes that the partition of the 
odd eight-vertex model can be rewritten
as follows\footnote{Here we also add that the partition function of the
odd eight-vertex model for periodic boundary conditions 
on $\mathrm{N} \times \mathrm{N}$ square lattice is always zero 
when the number of sites $\mathrm{N}$ is odd.},
\EQ
\label{ZOD}
\mathrm{Z}_{\mathrm{8\mathrm{od}}}(v_1,\dots,v_8)= 
\mathrm{Z}_{\mathrm{8{\mathrm{ev}}}}^{(\mathrm{stag})}\left (\substack{\underbrace{v_1,v_2,v_3,v_4,v_5,v_6,v_7,v_8} \\ \mathrm{X}} 
;\substack{\underbrace{v_3,v_4,v_1,v_2,v_8,v_7,v_6,v_5} \\ \mathrm{Y}} \right ), 
\EN
where the weights on the right hand side of Eq.(\ref{ZOD}) are organized
as in Figure (\ref{figure1}) with the identification $w_i=v_i$.

This equivalence was used to compute the free-energy of the odd eight-vertex model 
in the thermodynamic
limit \cite{WK} when the weights
satisfy the free-fermion condition, 
\EQ
\label{FREE}
v_1 v_2 +v_3 v_4 -v_5 v_6 -v_7 v_8=0,
\EN
but the possibility of building commuting transfer matrices for finite lattices 
either within the restriction (\ref{FREE})
or for any other manifold
has not yet been studied in the literature.

The same reasoning devised by Wu and Kunz \cite{WK} can be used 
to relate the partition function 
of the even eight-vertex model with that of 
an alternating odd eight-vertex model. In other
words one can also establish the following correspondence,
\EQ
\label{ZEV}
\mathrm{Z}_{\mathrm{8\mathrm{ev}}}(w_1,\dots,w_8)= 
\mathrm{Z}_{\mathrm{8{\mathrm{od}}}}^{(\mathrm{stag})}\left (\substack{\underbrace{w_1,w_2,w_3,w_4,w_5,w_6,w_7,w_8} \\ \mathrm{X}} 
;\substack{\underbrace{w_3,w_4,w_1,w_2,w_8,w_7,w_6,w_5} \\ \mathrm{Y}} \right ), 
\EN
where now the right hand side weights in Eq.(\ref{ZEV}) are read off
according to the Figure (\ref{figure2}) with $v_i=w_i$.

The above relationships become simpler when the weights of both models are symmetric. Taking into 
account the identifications (\ref{IDEN},\ref{IDEN1}) we obtain,
\EQ
\label{ZSYM1}
\mathrm{Z}_{\mathrm{8\mathrm{od}}}({\bf a},\dots,{\bf d})= 
\mathrm{Z}_{\mathrm{8{\mathrm{ev}}}}^{(\mathrm{stag})}\left (\substack{\underbrace{{\bf a},{\bf b},{\bf c},{\bf d}} \\ \mathrm{X}} 
;\substack{\underbrace{{\bf b},{\bf a},{\bf d},{\bf c}} \\ \mathrm{Y}} \right ), 
\EN
as well as,
\EQ
\label{ZSYM2}
\mathrm{Z}_{\mathrm{8\mathrm{ev}}}(a,\dots,d)= 
\mathrm{Z}_{\mathrm{8{\mathrm{od}}}}^{(\mathrm{stag})}\left (\substack{\underbrace{ a,b,c, d} \\ \mathrm{X}} 
;\substack{\underbrace{b,a,d,c} \\ \mathrm{Y}} \right ). 
\EN

We now note that the integrable manifold (\ref{MANI}) of the symmetric even eight-vertex
is invariant by the weights exchange $a \leftrightarrow b$ and $c \leftrightarrow d$. This fact
together with the equivalences (\ref{ZSYM1},\ref{ZSYM2}) suggest that
the odd eight-vertex model with symmetric weights 
may also have commuting transfer matrices on the manifold (\ref{MANI}). In next
section we show that this is indeed the case and discuss the respective Yang-Baxter
equation. We finally remark
that recently other
mappings among
the even and odd eight-vertex have been investigated by Assis \cite{ASS}. However,
they appear not useful for symmetric models since stringent conditions among
the weights are required.

\section{Yang-Baxter Relations}

In two spatial dimensions the method of commuting 
transfer matrices provides an efficient
device to uncover integrable lattice 
systems of statistical mechanics \cite{BAX}. For vertex models
the transfer matrix can be built by tensoring a number of local operators 
$\mathrm{L}_{\alpha,\beta}(\omega)$ which acts
on the product of two spaces associated with 
the horizontal ($\alpha$) and vertical ($\beta$)
edges statistical configurations. Here the symbol $\omega$ represents the set of
energy weights of the vertex model. Assuming periodic boundary conditions
on a square lattice of size 
$\mathrm{N} \times \mathrm{N}$ the transfer matrix can be represented
by the following trace, 
\begin{equation}
\label{TRA}
\mathrm{T}(\omega)=\mathrm{Tr}_0\left[ 
\mathrm{L}_{01}(\omega)
\mathrm{L}_{02}(\omega) \dots 
\mathrm{L}_{0\mathrm{N}}(\omega)\right], 
\end{equation}
where the index $0$ denotes the 
space associated to the horizontal degrees of freedom. 

In our specific case the horizontal and vertical spaces 
are two-dimensional
and the corresponding Lax operators can therefore be 
represented in terms of $4\times 4$ matrices. The form
of this matrix for the symmetric even eight-vertex models is, 
\EQ
\mathrm{L}^{(\mathrm{ev})}(a,b,c,d)=\left[
\begin{array}{cc|cc}
a & 0 & 0 & d \\
0 & b & c & 0 \\ \hline
0 & c & b & 0 \\
d & 0 & 0 & a \\
\end{array}
\right],
\EN
while for the symmetric odd eight-vertex model one has,
\EQ
\label{LODD}
\mathrm{L}^{(\mathrm{od})}({\bf a},{\bf b},{\bf c},{\bf d})=\left[
\begin{array}{cc|cc}
0 & {\bf a} & {\bf d} & 0 \\
{\bf b} & 0 & 0 & {\bf c} \\ \hline
{\bf c} & 0 & 0 & {\bf b} \\
0 & {\bf d}  & {\bf a} & 0 \\
\end{array}
\right].
\EN

It turns out that these operators can be related to each other by a weight independent
local transformation\footnote{Note that this does not corresponds to either a gauge or a twist transformation.}, namely
\EQ
\label{TRAN}
\mathrm{L}^{(\mathrm{od})}({\bf a},{\bf b},{\bf c},{\bf d})=\left(\sigma^{x} \otimes \sigma^{x} \right)
\mathrm{L}^{(\mathrm{ev})}({\bf a},{\bf b},{\bf c},{\bf d})\left(\sigma^{x} \otimes \mathrm{I}_2 \right),
\EN
where $\sigma^{x}$ is the standard Pauli matrix and $\mathrm{I}_2$ refers to the $2 \times 2$
identity matrix. Thanks to this correspondence we, from now on, can drop the use of bold letters
to distinguish the even from the odd energy weights.

An immediate consequence of the above transformation is 
that the transfer matrices 
of the even and odd
symmetric eight-vertex model are connected through a simple expression.
By substituting Eq.(\ref{TRAN}) in Eq.(\ref{TRA}) we see that 
the action of the matrices $\sigma^{x}$ on 
the auxiliary space are canceled out
and as result we obtain,
\EQ
\label{RELTRA}
\mathrm{T}^{(\mathrm{od})}(a,\dots,d)= 
\left(\sigma_1^{x} \otimes \sigma_2^{x} \otimes \dots \otimes \sigma_{\mathrm{N}}^{x} \right) \mathrm{T}^{(\mathrm{ev})}(a,\dots,d), 
\EN
where $\sigma_j^{x}$ acts as $\sigma^{x}$ on the j-$\mathrm{th}$ space and as identity otherwise. 
In addition, the symmetry under inversion of arrows orientations implies the following commutation relations,
\EQ
[\mathrm{T}^{(\mathrm{ev})}(a,\dots,d), 
\sigma_1^{x} \otimes \sigma_2^{x} \otimes \dots \otimes \sigma_{\mathrm{N}}^{x}]= 
[\mathrm{T}^{(\mathrm{od})}(a,\dots,d), 
\sigma_1^{x} \otimes \sigma_2^{x} \otimes \dots \otimes \sigma_{\mathrm{N}}^{x}]= 0.
\EN

An immediate consequence of these results is that 
both even and the odd eight-vertex models with symmetric weights have
in fact commuting transfer matrices when the weights are sited 
on the algebraic manifold (\ref{MANI}). 
The next natural step is to uncover
such  integrable manifold directly from the study of the 
Yang-Baxter equation
for the symmetric odd vertex model. To this end we need to 
find the form
of the $\mathrm{R}$-matrix satisfying the following relation,
\EQ
\label{YBAX}
\mathrm{R}_{12}(\omega^{'},\omega^{''})  
\mathrm{L}_{13}^{(\mathrm{od})}(a^{'},\dots,d^{'})
\mathrm{L}_{23}^{(\mathrm{od})}(a^{''},\dots,d^{''})=
\mathrm{L}_{23}^{(\mathrm{od})}(a^{''},\dots,d^{''})
\mathrm{L}_{13}^{(\mathrm{od})}(a^{'},\dots,d^{'})
\mathrm{R}_{12}(\omega^{'},\omega^{''}).  
\EN

In order to determine the $\mathrm{R}$-matrix we first take 
two distinct 
numerical points $a^{'},\dots,d^{'}$ and $a^{''},\dots,d^{''}$ on the curve (\ref{MANI}). We then solve
numerically the Yang-Baxter equation (\ref{YBAX}) for a general $16 \times 16$
$\mathrm{R}$-matrix and conclude that many matrix elements vanish. The
form of the $\mathrm{R}$-matrix is found to be similar to that
of the even eight-vertex model, namely
\EQ
\label{RPROP}
\mathrm{R}(\omega^{'},\omega^{''})=\left[
\begin{array}{cc|cc}
{\bf r_1} & 0 & 0 & {\bf r_4} \\
0 & {\bf r_2} & {\bf r_3} & 0 \\ \hline
0 & {\bf r_3} & {\bf r_2} & 0 \\
{\bf r_4} & 0 & 0 & {\bf r_1} \\
\end{array}
\right],
\EN
and therefore the $\mathrm{R}$-matrix and the Lax operator of the 
odd eight-vertex model
have distinct structures.

We next substitute the Lax operator (\ref{LODD}) and the $\mathrm{R}$-matrix 
proposal (\ref{RPROP}) in 
the Yang-Baxter equation and as a result we obtain 
six independent functional relations. Their explicit expressions are, 
\begin{eqnarray}
\label{YB1}
&& {\bf r_4}c^{'}b^{''} + {\bf r_1}a^{'}c^{''} - {\bf r_2}a^{'}d^{''} -{\bf r_3}c^{'}a^{''}=0, \nonumber \\
&& {\bf r_1}d^{'}a^{''} +{\bf r_4}b^{'}d^{''} -{\bf r_2}c^{'}a^{''} - {\bf r_3}a^{'}d^{''} =0, \nonumber \\
&& {\bf r_4}d^{'}a^{''} +{\bf r_1}b^{'}d^{''} - {\bf r_3}d^{'}b^{''} - {\bf r_2}b^{'}c^{''}=0, \nonumber \\
\label{YB4}
&& {\bf r_1}c^{'}b^{''} + {\bf r_4}a^{'}c^{''} -{\bf r_2}d^{'}b^{''}- {\bf r_3}b^{'}c^{''} =0, \nonumber \\
&& {\bf r_1}a^{'}b^{''} + {\bf r_4}c^{'}c^{''} -{\bf r_1}b^{'}a^{''} - {\bf r_4}d^{'}d^{''}=0, \nonumber \\
\label{YB6}
&& {\bf r_2}a^{'}a^{''} +{\bf r_3}c^{'}d^{''} - {\bf r_2}b^{'}b^{''} - {\bf r_3}d^{'}c^{''}=0 .
\end{eqnarray}

At this point we observe that such functional equations are similar to those associated with
the even eight-vertex model \cite{BAX}. They become exactly the same relations once we identify
the entries of the $\mathrm{R}$-matrix with the weights as follows,
\EQ
{\bf r_1} \rightarrow c,~~~
{\bf r_2} \rightarrow d,~~~
{\bf r_3} \rightarrow a,~~~
{\bf r_4} \rightarrow b,
\EN
and the invariants (\ref{MANI}) can therefore be derived along the lines already 
discussed by Baxter \cite{BAX}.

The above result suggests that the Yang-Baxter structure sitting on the curve (\ref{MANI}) is
richer than that solely associated to the even eight-vertex model. In order to see this
feature it is convenient to recall the weights parameterization introduced by Baxter \cite{BAX},
\begin{eqnarray}
\label{WEI}
&& a(\mu) = -\imath \Theta[\imath \lambda] \mathrm{H}\left[ \frac{\imath}{2}(\lambda-\mu) \right] \Theta \left[\frac{\imath}{2}(\lambda+\mu)\right], \nonumber \\
&& b(\mu) = -\imath \Theta[\imath \lambda] \Theta \left[ \frac{\imath}{2}(\lambda-\mu) \right] \mathrm{H} \left[\frac{\imath}{2}(\lambda+\mu)\right], \nonumber \\
&& c(\mu) = -\imath \mathrm{H}[\imath \lambda] \Theta \left[ \frac{\imath}{2}(\lambda-\mu) \right] \Theta \left[\frac{\imath}{2}(\lambda+\mu)\right], \nonumber \\
&& d(\mu) = \imath \mathrm{H}[\imath \lambda] \mathrm{H} \left[ \frac{\imath}{2}(\lambda-\mu)\right] \mathrm{H} \left[\frac{\imath}{2}(\lambda+\mu) \right],
\end{eqnarray}
where $\mu$ represents the curve spectral variable, $\lambda$ is a free parameter 
and $\mathrm{H}(\mu)$ and $\Theta(\mu)$ are theta
functions of modulus $k$ as defined in \cite{BAX}. 

We now define a family of $\mathrm{R}$-matrices as follows\footnote{Note that the index
exchange $\mathrm{ev} \leftrightarrow \mathrm{od}$ is equivalent to the weights 
replacements $a(\mu) \leftrightarrow c(\mu)$
and $b(\mu) \leftrightarrow d(\mu)$.},
\begin{eqnarray}
&& \mathrm{R}^{(\mathrm{ev},\mathrm{ev})}(\mu)=\left[
\begin{array}{cc|cc}
a(\mu) & 0 & 0 & d(\mu) \\
0 & b(\mu) & c(\mu) & 0 \\ \hline
0 & c(\mu) & b(\mu) & 0 \\
d(\mu) & 0 & 0 & a(\mu) \\
\end{array}
\right],~~~
\mathrm{R}^{(\mathrm{od},\mathrm{od})}(\mu)=\left[
\begin{array}{cc|cc}
c(\mu) & 0 & 0 & b(\mu) \\
0 & d(\mu) & a(\mu) & 0 \\ \hline
0 & a(\mu) & d(\mu) & 0 \\
b(\mu) & 0 & 0 & c(\mu) \\
\end{array}
\right], \nonumber \\
&& \mathrm{R}^{(\mathrm{od},\mathrm{ev})}(\mu)=\left[
\begin{array}{cc|cc}
0 & a(\mu) & d(\mu) & 0 \\
b(\mu) & 0 & 0 & c(\mu) \\ \hline
c(\mu) & 0 & 0 & b(\mu) \\
0 & d(\mu) & a(\mu) & 0 \\
\end{array}
\right],~~~
\mathrm{R}^{(\mathrm{ev},\mathrm{od})}(\mu)=\left[
\begin{array}{cc|cc}
0 & c(\mu) & b(\mu) & 0 \\
d(\mu) & 0 & 0 & a(\mu) \\ \hline
a(\mu) & 0 & 0 & d(\mu) \\
0 & b(\mu) & c(\mu) & 0 \\
\end{array}
\right]. \nonumber \\
\end{eqnarray}

By direct computations using the addition properties 
of teta functions one can verify
that these $\mathrm{R}$-matrices fulfill the following  
set of Yang-Baxter relations,
\EQ
\label{YBAXSH}
\mathrm{R}_{12}^{(\alpha_1,\alpha_2)}(\mu_1)  
\mathrm{R}_{13}^{(\alpha_1,\alpha_3)}(\mu_1+\mu_2)  
\mathrm{R}_{23}^{(\alpha_2,\alpha_3)}(\mu_2)=  
\mathrm{R}_{23}^{(\alpha_2,\alpha_3)}(\mu_2)  
\mathrm{R}_{13}^{(\alpha_1,\alpha_3)}(\mu_1+\mu_2)  
\mathrm{R}_{12}^{(\alpha_1,\alpha_2)}(\mu_1),  
\EN
where the upper indices $\alpha_j=\mathrm{ev}, \mathrm{od}$
giving rise to family
of Yang-Baxter relations. 
Note that the Yang-Baxter equation for the odd eight-vertex
model is selected with the choice $\alpha_1=\alpha_2=\mathrm{od}$ 
and $\alpha_3=\mathrm{ev}$.

Interesting enough, the above scenario is similar to 
that of solutions of the Yang-Baxter 
equation associated to different representations 
of a given group 
symmetry having the
same dimension.

\section{Concluding Remarks}

The even eight-vertex model is known to be solvable in the thermodynamic limit
when the weights satisfy the free-fermion condition \cite{FW}. This means that 
an exact expression for the infinity volume free-energy can be written
provided that,
\EQ
\label{FREE2}
w_1 w_2 +w_3 w_4 -w_5 w_6 -w_7 w_8=0.
\EN

However, in order to have commuting transfer matrices for finite
lattices sizes it is necessary the addition of other 
algebraic constraints. This has been first  pointed out
by Krinsky \cite{KR} when the weights satisfy the condition
$w_6=w_5$ and $w_8=w_7$. It is possible to adapt this work for
eight arbitrary weights and as a result one finds
the need of three extra restrictions on the weights. They are given by,
\EQ
\label{CONS2}
\frac{w_6 w_8}{w_5 w_7}=\Delta_1,~~~\frac{w_1 w_4+w_2 w_3}{w_5 w_7}=\Delta_2,~~~\frac{w_1^2+w_4^2-w_2^2-w_3^2}{w_5 w_7}=\Delta_3,
\EN
where $\Delta_1,\Delta_2$ and $\Delta_3$ are free parameters.

At this point we recall that Wu and Kunz have shown that 
free-energy of the odd eight-vertex model
can also be computed in the thermodynamic limit when the weights 
fulfill the 
free-fermion condition (\ref{FREE}). This is because the equivalent
staggered even eight-vertex model has also weights on
the free-fermion condition and the
infinite volume free-energy can again be computed by means 
of Pfaffian solution \cite{HLW}.
Motivated by this fact we attempt to check whether or not it is 
possible to construct
commuting transfer matrices for the odd eight-vertex model under 
the free-fermion condition (\ref{FREE}). 
By explicit computation 
of transfer matrices commutators already for 
two sites we conclude that if 
such manifold exists 
it is certainly not 
given by the constraints (\ref{CONS2}) when
the weight $w_i$ is replaced by $v_i$. This however does not exclude
the existence of another type of integrable manifold
for the asymmetric odd eight-vertex model with weights satisfying
the free-fermion condition. Otherwise the free-fermion 
odd eight-vertex model
will be an example in which the free-energy can be exactly computed
in the continuum without the existence of commuting transfer matrices
for finite lattice sizes. This issue appears to be worth of further
investigation.

We finally note that the mapping (\ref{ZEV}) suggests
that the staggered odd eight-vertex may
have commuting transfer matrices when the weights 
are lying on the manifolds (\ref{FREE2},\ref{CONS2}) with
$w_i=v_i$. In this case the partition function can be 
written as a trace
of a power of the product of two different operators, namely
\EQ
\mathrm{Z}_{\mathrm{8{\mathrm{od}}}}^{(\mathrm{stag})}(v_1,\dots,v_8)=
\mathrm{Tr}\left[ \left( \mathrm{T}_1(v_1,\dots,v_8) \mathrm{T}_2(v_1,\dots,v_8) \right)^{\mathrm{N}} \right],
\EN
where the corresponding transfer matrices are built by alternating two types of Lax operators,
\begin{eqnarray}
&& \mathrm{T}_1(v_1,\dots,v_8)=\mathrm{Tr}_0\left[ 
\mathrm{L}_{01}(v_1,\dots,v_8)
\mathrm{{\bf L}}_{02}(v_1,\dots,v_8) \dots 
\mathrm{L}_{02\mathrm{N}-1}(v_1,\dots,v_8)
\mathrm{{\bf L}}_{02\mathrm{N}}(v_1,\dots,v_8)\right], \nonumber \\
&& \mathrm{T}_2(v_1,\dots,v_8)=\mathrm{Tr}_0\left[ 
\mathrm{{\bf L}}_{01}(v_1,\dots,v_8)
\mathrm{L}_{02}(v_1,\dots,v_8) \dots 
\mathrm{{\bf L}}_{02\mathrm{N}-1}(v_1,\dots,v_8)
\mathrm{L}_{02\mathrm{N}}(v_1,\dots,v_8)\right]. \nonumber \\
\end{eqnarray}

The Lax operators encode the staggered 
weights $\omega_{\mathrm{X}}$ and 
$\omega_{\mathrm{Y}}$ as stated in the right-hand side of the correspondence (\ref{ZEV}) with
$w_i=v_i$. Their explicit matrices representation are,
\EQ
\label{LAXT}
\mathrm{L}(v_1,\dots,v_8)=\left[
\begin{array}{cc|cc}
0 & v_1 & v_7 & 0 \\
v_3 & 0 & 0 & v_6 \\ \hline
v_5 & 0 & 0 & v_4 \\
0 & v_8  & v_2 & 0 \\
\end{array}
\right],~~~
\mathrm{{\bf L}}(v_1,\dots,v_8)=\left[
\begin{array}{cc|cc}
0 & v_3 & v_6 & 0 \\
v_1 & 0 & 0 & v_7 \\ \hline
v_8 & 0 & 0 & v_2 \\
0 & v_5  & v_4 & 0 \\
\end{array}
\right].
\EN

We have verified for $\mathrm{N}=1,2$ 
that the operator product 
$\mathrm{T}_1(v_1,\dots,v_8)  
\mathrm{T}_2(v_1,\dots,v_8)$ commute for two distinct set of random numerical 
values for the weights sited on the manifolds (\ref{FREE2},\ref{CONS2}) with $w_i=v_i$.
It would be of interest to confirm this fact both 
algebraically and for arbitrary $\mathrm{N}$
considering the respective Yang-Baxter equation for the tensor product
of the two types of Lax operators (\ref{LAXT}).  Here we stress that 
the transfer matrices 
$\mathrm{T}_1(v_1,\dots,v_8)$ and 
$\mathrm{T}_2(v_1,\dots,v_8)$ 
do not have vanishing commutators for
distinct weights on the manifolds (\ref{FREE2},\ref{CONS2}). This appears to be an
example in which individual transfer matrices may not commute but the product
of two of them can lead us to a family of commuting operators.

\section*{Acknowledgments}

This work was supported in part by the Brazilian Research Council CNPq-2016/401694.

\end{document}